\documentclass[letterpaper,12pt]{article}
\usepackage{tabularx} 
\usepackage{amsmath}  
\usepackage{amssymb}
\usepackage[pdftex]{graphicx}
\pdfoutput=1
\usepackage[margin=1in,letterpaper]{geometry} 
\usepackage{cite} 
\usepackage[final]{hyperref} 
\hypersetup{
	colorlinks=true,       
	linkcolor=blue,        
	citecolor=blue,        
	filecolor=magenta,     
	urlcolor=blue
}
\newcommand{\Slash}[1]{{\ooalign{\hfil/\hfil\crcr$#1$}}}
\newcommand{\beq}{\begin{equation}}
\newcommand{\eeq}{\end{equation}}
\newcommand{\bea}{\begin{eqnarray}}
\newcommand{\eea}{\end{eqnarray}}

\newcommand{\ba}{\begin{array}}
\newcommand{\ea}{\end{array}}
\newcommand{\bla}{\left<}
\newcommand{\ket}{\right>}

\newcommand{\C}[1]{{\cal{#1}}}

\begin{document}

\setlength{\baselineskip}{0.7cm}
\begin{titlepage}
    \begin{flushright}
        NITEP 135 \\
        KEK-TH-2419 \\
        KEK-Cosmo-0295
    \end{flushright}
    \vspace{10mm}
    \begin{center}
        \Large\textbf{Fermion Mass Hierarchy and Mixing} \\
        \vspace{2mm}
        \Large\textbf{in simplified Grand Gauge-Higgs Unification} \\
    \end{center}
    \vspace{10mm}
    \begin{center}
        {\large Nobuhito Maru},$^{a,\,b}$
        {\large Haruki Takahashi}$^{\,c,\,d}$ and
        {\large Yoshiki Yatagai}$^{\,a}$
    \end{center}
    \vspace{0.2cm}
    \begin{center}
        ${}^{a}$ \textit{Department of Physics, 
        Osaka Metropolitan University, \\
        Osaka 558-8585, Japan} \\
        ${}^{b}$ \textit{Nambu Yoichiro Institute of Theoretical and Experimental Physics (NITEP), \\
        Osaka Metropolitan University, Osaka 558-8585, Japan} \\
        ${}^{c}$ \textit{KEK Theory Center, High Energy Accelerator Research Organization (KEK), \\
        Oho 1-1, Tsukuba, Ibaraki 305-0801, Japan} \\
        ${}^{d}$ \textit{The Graduate University for Advance Studies (SOKENDAI), \\
        Oho 1-1, Tsukuba, Ibaraki 305-0801, Japan} \\
    \end{center}
    \date{}
\vspace*{5mm}
    \begin{abstract}
Grand gauge-Higgs unification of five dimensional SU(6) gauge theory on an orbifold $S^1 / Z_2$
   with localized gauge kinetic terms is discussed.
The Standard model (SM) fermions on the boundaries and
   some massive bulk fermions coupling to the SM fermions 
   are introduced.
The number of the bulk fermions is reduced compared to the previous model, 
 which reproduces the generation mixing of the SM fermions 
in addition to the SM fermion mass hierarchy by mild tuning the bulk masses
 and parameters of the localized gauge kinetic terms.
    \end{abstract}
\end{titlepage}
\section{Introduction}
Gauge-Higgs unification (GHU) \cite{GHU} is one of the physics beyond the Standard Model (SM),
	which solves the hierarchy problem by identifying the SM Higgs field
	with one of the extra spatial component of the higher dimensional gauge field.
In this scenario, 
	the physical observables in Higgs sector are calculable and predictable regardless of its non-renormalizability.
For instance,
	the quantum corrections to Higgs mass and Higgs potential are known to be finite
	at one-loop \cite{1loop} and two-loop \cite{2loop} thanks to the higher dimensional gauge symmetry.

The hierarchy problem originally exists in grand unified theory (GUT)
 whether the discrepancy between the GUT scale and the week scale are kept
	and stable under quantum corrections.
Therefore, the extension of GHU to grand unification is a natural direction to explore.
One of the authors discussed a grand gauge-Higgs unification (GGHU) \cite{LM},
	\footnote{For earlier attempts and related works, see \cite{others}}
	where the five dimensional $SU(6)$ GGHU was considered
	and the SM fermions were embedded into zero modes of $SU(6)$ multiplets in the bulk.
This setup was very attractive because of the minimal matter content
	without massless exotic fermions absent in SM, namely an anomaly-free matter content.
However, a crucial drawback was found.
The down-type Yukawa couplings and the charged lepton Yukawa couplings in GHU
 originated from the gauge interaction cannot be allowed
	since the left-handed $SU(2)_L$ doublets
	and the right-handed $SU(2)_L$ singlets are embedded into different $SU(6)$ multiplets.
This fact seems to be generic in GHU as long as the SM fermions are embedded into the bulk fermions.
Fortunately, alternative approach to generate Yukawa coupling in a context of GHU has been known \cite{SSS, CGM},
 in which the SM fermions are introduced on the boundaries (i.e. fixed point in an orbifold compactification).
We also introduce massive bulk fermions,
	which couple to the SM fermions through the mass terms on the boundary.
Integrating out these massive bulk fermions leads to non-local SM fermion masses,
	which are proportional to the bulk to boundary couplings and exponentially sensitive to their bulk masses.
Then, the SM fermion mass hierarchy can be obtained by very mild tuning of bulk masses.

Along this line, we have improved an $SU(6)$ grand GHU model of \cite{LM} in \cite{MY1},
	where the SM fermion mass hierarchy except for top quark mass was obtained
	by introducing on the boundary as $SU(5)$ multiplets,
	the four types of massive bulk fermions in $SU(6)$ multiplets coupling to the SM fermions.
Furthermore, we have shown that the electroweak symmetry breaking and an observed Higgs mass can be realized
	by introducing additional bulk fermions with large dimensional representation.
In GHU, generation of top quark mass is difficult
	since Yukawa coupling is originally gauge coupling and fermion mass is at most an order of W boson mass as it stands.
As a useful approach \cite{SSS},
	introducing the localized gauge kinetic terms on the boundary is known to have enhancement effects on fermion masses.
In our previous paper \cite{MY2},
	we followed this approach in order to realize the SM fermion mass hierarchy including top quark.
Once the localized gauge kinetic terms are introduced,
	the zero mode wave functions of gauge fields are distorted and the gauge coupling universality is not guaranteed.
We found a parameter space where the gauge coupling constant between fermions and a gauge field,
	the cubic and the quartic self-coupling constants are almost universal.
Then, we showed that the fermion mass hierarchy including top quark mass was indeed realized
	by appropriately choosing the bulk mass parameters
	and the size of the localized gauge kinetic terms.
The correct pattern of electroweak symmetry breaking was obtained
	by introducing extra bulk fermions as in our paper \cite{MY2},
	but their representations have been greatly simplified.

However, the generation mixing could not be generated unfortunately in the previous model \cite{MY2} because each type of bulk fermions was introduced per generation. We have to modify our model in order to reproduce the generation mixing in addition to the SM fermion masses. As we will see later, this modification can be achieved by changing how the bulk fermions couple to the SM fermions on the boundaries and hence reducing the number of bulk fermions. This reduction makes reproducing the quark and lepton mixing angles highly nontrivial but finally enables us to reproduce the flavor mixing angles and a CP phase. After the analysis, we find allowed parameter sets to reproduce the SM fermion masses and mixing.


This paper is organized as follows.
In the next section, we briefly describe the gauge and Higgs sectors of our model with the localized gauge kinetic terms, and discuss the mass spectrum of gauge fields including their effects.
In section 3, we explain how our model has been changed
	and the generation mechanism of the SM fermion masses and mixing. 
Then, it is shown that the SM fermion masses and mixing can be reproduced
	by mild tuning of bulk masses and parameters of the localized gauge kinetic terms.
Final section is devoted to our conclusions.
\section{Setup}
In this section, we briefly explain our model \cite{MY2}, 
 which is
a five dimensional (5D) $SU(6)$ gauge theory compactified on an orbifold $S^1/Z_2$
 where the radius is denoted as $R$.
Since two fixed points are located at $y=0, \pi R$ in the fifth dimension, 
 $Z_2$ parities must be imposed and given as follows.
    \begin{eqnarray}
        P  &=& \mbox{diag}(+,+,+,+,+,-) \, \, \mbox{at}~y=0, \nonumber \\
        P' &=& \mbox{diag}(+,+,-,-,-,-) \, \, \mbox{at}~y=\pi R.
    \end{eqnarray}
Accordingly, we assign the $Z_2$ parity for the gauge field and the scalar field 
 as $A_{\mu}(-y)=P A_{\mu}(y) P^{\dag}$, $A_y(-y)=-P A_y(y) P^{\dag}$. 
We note that only the field with the parity $(+,+)$ has a 4D massless zero mode, 
	where $(+, +)$ means that $Z_2$ parity is even both at $y=0~(y=\pi R)$ boundary.
This is because the wave functions of $(+,+)$ are $\cos(n y/R)~(n=0, \cdots, \infty)$ 
 after the Kaluza-Klein (KK) expansion.
The $Z_2$ parity of $A_{\mu}$ tells us 
 that $SU(6)$ gauge symmetry is broken to $SU(3)_C \times SU(2)_L \times U(1)_Y \times U(1)_X$
	by the combination of symmetry breaking pattern at each boundary,
    \begin{eqnarray}
        && SU(6) \rightarrow SU(5)\times U(1)_X \hspace{1.6cm} \mbox{at}~y=0, \\
        && SU(6) \rightarrow SU(2)\times SU(4) \times U(1)' ~\mbox{at}~y=\pi R.
    \end{eqnarray}
The hypercharge $U(1)_Y$ is contained in Georgi-Glashow $SU(5)$ GUT,
which implies that the weak mixing angle is 
	$\sin^2{\theta_W}=3/8$ ($\theta_W$ :weak mixing angle) at the unification scale.

The problem of remaining extra $U(1)_X$ is easily resolved 
 by introducing a 4D $U(1)_X$ charged scalar field localized on a fixed point
 and constructing the quadratic and quartic scalar potential with negative mass squared.
The scalar field will develop a vacuum expectation value (VEV) 
 and the $U(1)_X$ gauge field is massive.

The SM $SU(2)_L$ Higgs doublet field is identified 
 with a part of an extra component of gauge field $A_y$. 
The vacuum expectation value (VEV) of the Higgs field is assumed to be taken in the 28-th $SU(6)$ generator as
 	$\langle A_y^a \rangle = \frac{2\alpha}{Rg}\delta^{a\,28}$,
	where $g$ is a 5D $SU(6)$ gauge coupling constant and $\alpha$ is a dimensionless constant.
The VEV of the Higgs field is expressed by $\langle H \rangle = \frac{\sqrt{2}\alpha}{Rg}$.
In this setup, the doublet-triplet splitting problem is solved
	by the orbifolding since the the colored Higgs has a $Z_2$ parity $(+,-)$
	and becomes massive correspondingly \cite{Kawamura}.

The Higgs couplings to the gauge bosons and the fermions are included 
 in the gauge interactions of their kinetic terms,
    \begin{equation}
        -\frac{1}{4} \mathcal{F}_{MN}^a \mathcal{F}^{a\,MN} \supset 
        -\frac{1}{2} \mathcal{F}_{\mu y}^a \mathcal{F}^{a\,\mu y} \supset 
        -\frac{1}{2} A_y^a (\partial_y + f^{adb} A_y^d)(\partial_y + f^{bec}A_y^e)A^{\mu\,c},
    \label{Equation8}
    \end{equation}
    \begin{equation}
        \overline{\Psi}i\Slash{D}\Psi \supset \overline{\Psi}iD_y\Gamma^y\Psi = -\overline{\Psi}(\partial_y+A_y)\gamma^5\Psi,
    \label{Equation9}
    \end{equation}
	where $M$, $N=\{\mu,y\}$, $\mu =0,1,2,3$, $y=5$
	and subscript $a$, $b$, $c$, $d$, $e$ denote the gauge indices for $SU(6)$.
$\Gamma^y$ in (\ref{Equation9}) is the fifth component of the five-dimensional gamma matrices
 $\Gamma^M=(\Gamma^{\mu}, \Gamma^y)=(\gamma^{\mu},i\gamma^5)$.
Eqs.\,(\ref{Equation8}) and \,(\ref{Equation9}) become the mass terms after the Higgs field takes the VEV. 
The mass eigenvalues are obtained as $m_n(q\alpha)=\frac{n+\nu+q\alpha}{R}$, 
 where $n$ is KK mode, $\nu=0$ or $1/2$.
$q$ is an integer charge fixed by the $SU(2)$ representation 
 to which the field coupled to Higgs field belongs.
If the field belongs to $\textbf{N+1}$ representation of $SU(2)_L$, the $q$ is equal to $N$.
For instance, $\textbf{6}$ representation of $SU(6)$ has the branching rule 
 under $SU(6)\rightarrow SU(3)_C\times SU(2)_L$ 
 like $\textbf{6}\rightarrow (3,1)\oplus (1,2)\oplus (1,1)$, 
which implies that this representation has four states with $q=0$ and one state with $q=1$.

In order to reproduce top quark mass, 
	we introduce additional gauge kinetic terms localized at $y=0$ and $y=\pi R$.
Lagrangian of $SU(6)$ gauge kinetic term is
    \begin{equation}
        \mathcal{L}_g = -\frac{1}{4} \mathcal{F}^{a\,MN}\mathcal{F}^a_{MN}
        -2\pi Rc_1\delta(y)\frac{1}{4}\mathcal{F}^{b\,\mu\nu}\mathcal{F}^b_{\mu\nu}
        -2\pi Rc_2\delta (y-\pi R)\frac{1}{4}\mathcal{F}^{c\,\mu\nu}\mathcal{F}^c_{\mu\nu},
    \label{Equation10}
    \end{equation}
	where the first term is the 5D bulk gauge kinetic term, 
 the second and the third terms are gauge kinetic terms 
  localized at fixed points.
$c_{1,2}$ are dimensionless free parameters.
The superscripts $a,b,c$ denote the gauge indices for $SU(6)$, $SU(5)\times U(1)$, $SU(2)\times SU(4)\times U(1)'$. 

The mass spectrum of the SM gauge field becomes very complicated by these localized gauge kinetic terms. 
In particular, their effects for a periodic sector where $A(y+\pi R)=A(y)$ or the fields with parity $(P,P^{'})=(+,+),(-,-)$
and an anti-periodic where $A(y+\pi R)=-A(y)$ or the fields with parity $(P,P^{'})=(+,-),(-,+)$
sector are different. 
In a basis where 4D gauge kinetic terms are diagonal,
	the boundary conditions for wave functions are found 
	as $f_n(y+\pi R;q\alpha)=e^{2i\pi q\alpha}f_n(y;q\alpha)$ in periodic sector
	and $f_n(y+\pi R;q\alpha)=e^{2i\pi(q\alpha +1/2)}f_n(y;q\alpha)$ in anti-periodic sector.
The wave functions in the same basis satisfy the following equation
    \begin{equation}
        \left[\partial_y^2+m_n^2(q\alpha) \left(1+2\pi Rc_1\delta(y)+2\pi Rc_2\delta(y-\pi R)\right)\right]f_n(y;q\alpha)=0,
    \label{equation11}
    \end{equation}
	where $m_n(q\alpha)$ is the KK mass.
By solving eq.\,(\ref{equation11}) with the (anti-)periodic 
 boundary conditions,
	the equations determining the KK mass spectrum are obtained \cite{CTW}.
    \begin{equation}
        2(1-c_1 c_2\xi_n^2)\sin^2\xi_n+(c_1+c_2)\xi_n\sin 2\xi_n -2\sin^2(\pi(q\alpha+\nu))=0
    \end{equation}
where $\xi_n=\pi Rm_n$ and $\nu$ is $0~(1/2)$ for the periodic (anti-periodic) sector.

Taking into account that $m_0$ is around weak scale ($\sim 100$ GeV) and $1/R$ is larger than 1 TeV,
	it is reasonable to suppose $\xi_0 \ll 1$.
Then, we can find an approximate form of $\xi_0$ as
    \begin{equation}
        \xi_0 \sim \frac{\sin(\pi(q\alpha+\nu))}{\sqrt{1+c_1+c_2}}.
    \end{equation}
For instance, 
 the W boson mass $m_W$ is given by
    \begin{equation}
        m_W=\frac{\sin(\pi \alpha)}{\pi R\sqrt{1+c_1+c_2}} 
    \end{equation}
  since the W boson is the gauge boson with $q=1$ and $\nu=0$  
    
\section{Fermion masses and mixing}
In the previous paper \cite{MY2},
	the SM fermions were embedded into $SU(5)$ multiplets localized at $y=0$ boundary,
	where three sets of decouplet, anti-quintet and singlet $\chi_{10},\, \chi_{5^*},\, \chi_1$ are introduced.
We also introduced three types of bulk fermions $\Psi$ and $\Tilde{\Psi}$ (referred as ``mirror fermions")
	with opposite $Z_2$ parities each other per a generation and constant mass term such as $M\Bar{\Psi}\Tilde{\Psi}$ in the bulk
	to avoid exotic 4D massless fermions.
Without these mirror fermions and mass terms,
 	we necessarily have extra exotic 4D massless fermions with the SM charges after an orbifold compactification.
As a result, we have no massless chiral fermions from the bulk and its mirror fermions.
The massless fermions are only the SM fermions
	and the gauge anomalies for the SM gauge groups are trivially canceled.

In this setup, we could not generate the generation mixings of quarks and leptons \cite{MY2}. 
In this chapter, we will see how the model has been changed, 
 discuss the mechanism generating the SM mass and the generation mixing of weak interaction, 
 and then show the setup of our model and the results.

\subsection{Boundary fermion mass}
\begin{figure}[t]
			\begin{tabular}{cc}
				\begin{minipage}[c]{0.47\hsize}
					\centering
					\includegraphics[keepaspectratio,
														scale=0.7,
														]{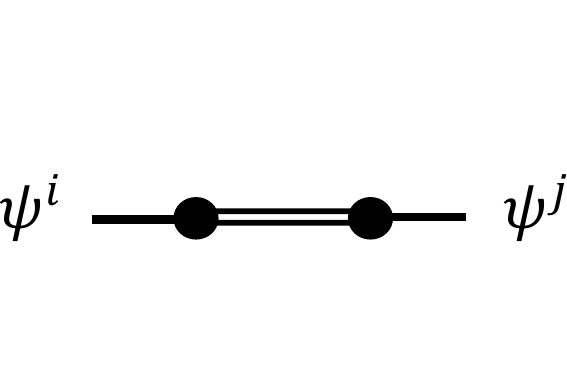}
					\caption{The diagram generating kinetic mixing of SM fermion $\psi^{i}$. 
										The double line represents the propagator of the bulk and mirror fermion.}
					\label{Figurekinetic}
				\end{minipage}
				&
				\begin{minipage}[c]{0.47\hsize}
					\centering
					\includegraphics[keepaspectratio,
														scale=0.7,
														]{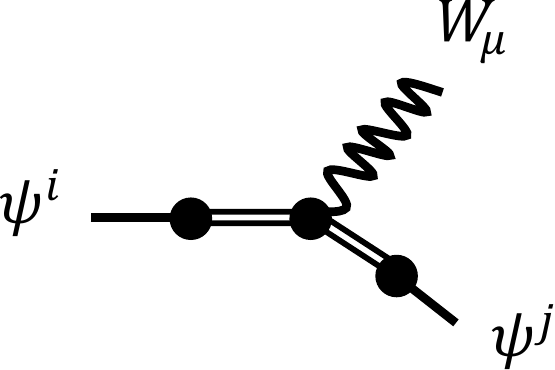}
					\caption{The diagram generating weak interaction of SM fermion.
										$\psi^{i}$ and $W$ are the SM fermion and the weak boson, respectively.
										The double line represents the propagate of the bulk and mirror fermion.}
					\label{Figureweak}
				\end{minipage}
			\end{tabular}
		\end{figure}

The boundary fermions $\psi^{i=1,2,3}$ are localized on the boundary, $y=0$ or $y=\pi R$,
	and these have the kinetic mixing terms between the bulk and mirror fermion.
These interaction terms are
	\begin{equation}
  	\begin{aligned}
    	\C{L}_4
      	&= \int dy
				 					\sum_{i}
									\sqrt{\frac{2}{\pi R}}
                	\left\{
                	\delta(y-r_1^i)
                 	\epsilon_L^i \overline{\psi_L^i}(x) A(x,y)
                	+\delta(y-r_2^i)
                 	\epsilon_R^i \overline{\psi_R^i}(x) B(x,y)
                	+\mbox{h.c.}\right\}
                 	\\
     		&=\sum_{i}
					\sum_n \frac{1}{\pi R}
              	\left\{
                  	\xi_{r_1^i}^a \epsilon_L^i \overline{\psi_L^i}(x) a_n(x)
                  	+ \xi_{r_2^i}^b \epsilon_R^i \overline{\psi_R^i}(x) b_n(x)
                  	+ \mbox{h.c.}
              	\right\},
  	\end{aligned}
	\end{equation}
	where $A$, $B$ are the bulk fermion or mirror fermion and
	$a$ and $b$ are the corresponding 4D fields:
	\begin{equation}
		A(x,y) = \sum_{n} f_n(y) a(x),
		\hspace{20pt}
		B(x,y) = \sum_{n} f_n(y) b(x),
	\end{equation}
	and $\xi_{r_n^i} =\sqrt{\pi R} f_n(x,r_n^i)= 1$ or $(-1)^n$.
 The generation mixing of the boundary fermion is generated
  by integrating out the bulk or mirror fermions, which can be seen from the diagram shown in Fig.~\ref{Figurekinetic},
	\begin{equation}
  	\begin{aligned}
				& \sum_{i,j}
						\left[
						-i\overline{\psi_L^i} \Slash{p}_E Z_L^{ij} \psi_L^j
          	-i\overline{\psi_R^i} \Slash{p}_E Z_R^{ij} \psi_R^j
          	+\left(\overline{\psi_L^i} M^{ij} \psi_R^j + \mbox{h.c.}\right)
						\right],\\
  	\end{aligned}
	\end{equation}
	where the subscript $E$ means Euclidean, namely Wick rotation is performed.
Furthermore, all terms can be expressed in terms of the bulk and mirror fermion propagator $\bla ... \ket_E$
	\begin{equation}
  	\begin{aligned}
    	\Slash{p}_E Z_L^{ij} P_L
				&= \frac{-i}{\pi^2 R^2}
						\sum_n \xi_{r_1^i}^a\xi_{r_2^j}^b
						\epsilon_L^i \epsilon_L^{j\ast}
						P_R \bla a_n(x)\overline{a}_n(x)\ket_E P_L,
						\\
    	\Slash{p}_E Z_R^{ij} P_R
				&= \frac{-i}{\pi^2 R^2}
						\sum_n \xi_{r_1^i}^a\xi_{r_2^j}^b
						\epsilon_R^i \epsilon_R^{j\ast}
						P_L\bla b_n(x)\overline{b}_n(x)\ket_E P_R,
						\\
    	M^{ij} P_R
				&= \frac{1}{\pi^2 R^2}
						\sum_n  \xi_{r_1^i}^a\xi_{r_2^j}^b
						\epsilon_L^i \epsilon_R^{j\ast}
						P_R \bla a_n(x)\overline{b}_n(x)\ket_E P_R.  \\
    	\label{eq1}
  	\end{aligned}
	\end{equation}
We notice that these mixing effects are new bulk contributions,
 which are absent in the previous paper \cite{MY2}
 where the same set of bulk and mirror fermions are introduced per generation.
By construction, we have no mixing from the bulk sector.
In this paper, however, the generation mixings are inevitable due to the reduction of the number of bulk and mirror fermions,
 which makes our analysis very complicated and also obtaining realistic mixing parameters nontrivial.

We can evaluate these terms (\ref{eq1})
 by computing the bulk and mirror fermion propagator $\bla ... \ket_E$ and the summation of the KK-modes.
The methods are summarized in appendix A and B.
The results are classified into three cases depending on the cases
	where $a$ and $b$ are the bulk fermion or the mirror fermion,
	and can be rewritten by the following functions,
	\begin{equation}
	  \begin{aligned}
	    f_{\delta}^{(T)}(x ,q\alpha)
	    &=\left\{
	    \begin{array}{ll}
	      \coth(x +i \pi \alpha) & :\delta=0, T=+1, (\nu_T=0)\\
	      \sinh(x +i \pi \alpha)^{-1} & :\delta=1, T=+1, (\nu_T=0)\\
	      \tanh(x +i \pi \alpha) & :\delta=0, T=-1, (\nu_T=1/2)\\
	      \cosh(x +i \pi \alpha)^{-1} & :\delta=-1, T=-1, (\nu_T=1/2)\\
	    \end{array}\right. ,
	  \end{aligned}
	\end{equation}
	where $x= \pi R p_E$
	and $\delta= 0$ ($r_1^i=r_2^j$) or $\delta= 1$ ($r_1^i \neq r_2^j$).

The first case is that the both are the bulk fermions:
	\begin{equation}
	  \begin{aligned}
	    Z_L^{ij} &= \frac{\epsilon_L^i \epsilon_L^{j\ast}}{\sqrt{x^2+\lambda^2}}
	    					\mbox{Re} f_{\delta}^{(T)} \left(\sqrt{x^2+\lambda^2},q\alpha \right),\\
	    Z_R^{ij} &= \frac{\epsilon_R^i \epsilon_R^{j\ast}}{\sqrt{x^2+\lambda^2}}
	    					\mbox{Re} f_{\delta}^{(T)} \left(\sqrt{x^2+\lambda^2},q\alpha \right),\\
	    M^{ij} &= \frac{\epsilon_L^i \epsilon_R^{j\ast}}{\pi R}
							\mbox{Im} f_{\delta}^{(T)} \left(\sqrt{x^2+\lambda^2},q\alpha \right),\\
	  \end{aligned}
	\end{equation}
	where $\lambda = \pi R M$.

The second case is that these are the bulk and mirror fermions, respectively:
	\begin{equation}
	  \begin{aligned}
			Z_L^{ij} &= 0,\\
	    Z_R^{ij} &= 0,\\
	    M^{ij} &= -\frac{\epsilon_L^i \epsilon_R^{j\ast} \lambda}{\pi R \sqrt{x^2+\lambda^2}}
							\mbox{Re} f_{\delta}^{(T)} \left(\sqrt{x^2+\lambda^2},q\alpha \right).\\
	  \end{aligned}
	\end{equation}
The third case is that the both are the mirror fermions:
	\begin{equation}
	  \begin{aligned}
			Z_L^{ij} &= \frac{\epsilon_L^i \epsilon_L^{j\ast}}{\sqrt{x^2+\lambda^2}}
								\mbox{Re} f_{\delta}^{(T)} \left(\sqrt{x^2+\lambda^2},q\alpha \right),\\
	    Z_R^{ij} &= \frac{\epsilon_R^i \epsilon_R^{j\ast}}{\sqrt{x^2+\lambda^2}}
								\mbox{Re} f_{\delta}^{(T)} \left(\sqrt{x^2+\lambda^2},q\alpha \right),\\
	    M^{ij} &= -\frac{\epsilon_L^i \epsilon_R^{j\ast}}{\pi R}
								\mbox{Im} f_{\delta}^{(T)} \left(\sqrt{x^2+\lambda^2},q\alpha \right).
								\label{eqmassM}\\
	  \end{aligned}
	\end{equation}

If some bulk and mirror fermions are introduced,
	all of the contributions for the kinetic and mass mixing must be summed in eq.\,(\ref{eq1}),
	\begin{eqnarray}
	    \tilde{Z}_L^{ij}
							&\equiv& \delta^{ij} +\sum_{a} Z_{L}^{a,ij},\\
	    \tilde{Z}^R_{ij}
							&\equiv& \delta^{ij} + \sum_{a} Z_{R}^{a,ij},\\
	    \tilde{M}^{ij}
							&\equiv& \sum_{a} M^{a,ij} .
	\end{eqnarray}
In the expression above,
	the superscript $``a"$ in $Z_{L(R)}^{a,ij}$ and $M^{a,ij}$
	shows the contributions from some bulk and mirror fermions.
	.

These kinetic mixing can be diagonalized by unitary matrices $U_{Z_{L,R}}$
	\begin{eqnarray}
	  \tilde{Z}_L = U_{Z_L}^{\dag} Z_L^{\textmd{diag}} U_{Z_L}, \\
	  \tilde{Z}_R = U_{Z_R}^{\dag} Z_R^{\textmd{diag}} U_{Z_R},
	\end{eqnarray}
	where $Z_{L,R}^{\textmd{diag}}$ are diagonal matrices.
After rewriting the mass term
	in terms of new basis  diagonalizing the kinetic mixings and normalizing the kinetic term,
	we obtain the following mass matrix
	\begin{equation}
	   M' = \sqrt{(Z_L^{\textmd{diag}})^{-1}}
					U_{Z_L}
					\tilde{M}
					U_{Z_R}^{\dag}
					\sqrt{(Z_R^{\textmd{diag}})^{-1}}.
	\end{equation}
Next, we perform unitary transformations in order to move on to the mass basis:
	\begin{equation}
	     \psi'_R = U_R \psi_R,\hspace{30pt} \psi'_L = U_L \psi_L.
			 \label{eqnewbase}
	\end{equation}
In this new basis, the mass matrix is diagonalized:
	\begin{equation}
	     M' = U^{\dag}_L M_{\textmd{diag}} U_R.
	\end{equation}
This expression allows us to compute the SM fermion masses.

\subsection{Weak interaction for boundary fermion}
As in the previous section, taking into account the mixing effects
	between the bulk and mirror fermions and the boundary fermions,
	the additional mixing in the charged weak interaction of the boundary fermion,
	which is not present in the SM,
       also seems to be generated.
The corresponding interactions generating the additional mixings in the charged weak interaction are shown below,
	\begin{equation}
  	\begin{aligned}
    	\C{L}_4
     	\supset
      	& \sum_{i}\left[
						W^{+}_{\mu}\overline{u_L^i}\gamma^{\mu}d_L^i
						+\sum_n \frac{1}{\pi R}
		              	\epsilon_L^i
										\left(
		                  	\xi_{r_1^i}^u \overline{u_L} a^u_n
		                  	+ \xi_{r_2^i}^d \overline{d_L} a^d_n
		              	\right)\right]
           	\\
      	&+\sum_n W^{+}_{\mu}\overline{\chi^u_n}\gamma^{\mu}\chi^d_n
				+\sum_n W^{+}_{\mu}\overline{\tilde{\chi}^u_n}\gamma^{\mu}\tilde{\chi}^d_n
				+ \mbox{h.c.},
  	\end{aligned}
	\end{equation}
	where $W$ is the weak boson,
	$u$ is up-type quark or electron,
	$d$ is down-type quark or neutrino,
	$a$ and $b$ are the bulk or mirror fermion,
	$\chi$ and $\tilde{\chi}$ are the bulk and mirror fermion, respectively.
The additional generation mixing in the charged weak interaction is generated
    by integrating out the bulk and mirror fermions
     and can be seen from the diagram shown in Fig.~\ref{Figureweak},
	\begin{equation}
  	\begin{aligned}
						 \sum_{ij} W^{+}_{\mu} \overline{u_L^i}\gamma^{\mu} M_{wi}^{ij} d_L^j +\mbox{h.c.},\\
  	\end{aligned}
	\end{equation}
	where
	\begin{equation}
  	\begin{aligned}
    	\gamma^{\mu}M_{wi}^{ij}P_L
    		&=P_R  \sum_n
           	\frac{\xi_{r_1^i}^u  \xi_{r_2^j}^d
           	\epsilon_L^i \epsilon_L^{j\ast}}{(\pi R)^2}
           	\left\{
           	\bla a^u_n \overline{\chi^u_n} \ket
           	\gamma^{\mu}
           	\bla \overline{a^d_n} \chi^d_n \ket
         +\bla a^u_n \overline{\tilde{\chi}^u_n} \ket
           	\gamma^{\mu}
           	\bla \overline{a^d_n} \tilde{\chi}^d_n \ket
           	\right\} P_L.
  	\end{aligned}
		\label{eqw}
	\end{equation}
Similary, this contribution can be evaluated by using the method in appendix A and B.
Then, we found the additional mixings in the charged weak interaction
 to be the same as the kinetic mixing for the left-handed fermions localized on the boundary,
	\begin{equation}
		\begin{aligned}
			M_{wi}^{ij} = Z_L^{ij}.
 		\end{aligned}
	\end{equation}
This property ensures that no additional mixings is needed as will be shown below.
In the case where some bulk and mirror fermions are introduced,
	the kinetic and mass mixing are the summation of all contributions in eq.\,(\ref{eqw}),
 \begin{equation}
 		\tilde{M}_{wi}^{ij} = \delta^{ij} +\sum_a M_{wi,a}^{ij}
												= \tilde{Z}_L^{ij}.
 \end{equation}
After rewriting the charged weak interaction in terms of new basis in eq.~(\ref{eqnewbase}),
 	CKM matrix and PMNS matrix are given by
	\begin{equation}
		\begin{aligned}
			V_{\textmd{CKM}}
				&= U^u_L
				 		\sqrt{(Z_L^{\textmd{diag},u})^{-1}}
						U_{Z_L}^u
						\tilde{M}_{wi}^{ud}
						U_{Z_L}^{d\dag}
						\sqrt{(Z_L^{\textmd{diag},d})^{-1}}
						U^{d\dag}_L
				 = U^u_L U^{d\dag}_L ,\\
 			V_{\textmd{PMNS}}
				&= U^e_L
						\sqrt{(Z_L^{\textmd{diag},e})^{-1}}
						U_{Z_L}^e
						\tilde{M}_{wi}^{e\nu}
						U_{Z_L}^{\nu\dag}
						\sqrt{(Z_L^{\textmd{diag},\nu})^{-1}}
						U^{\nu\dag}_L
				 = U^e_L U^{\nu\dag}_L,
		\end{aligned}
	\end{equation}
where $u,d,e,\nu$ denote the up-type quarks, down-type quarks,
 charged leptons, and neutrinos of the SM fermion, respectively.
We utilized the fact that the contributions of the left-handed $SU(2)$ doublets to the kinetic mixing are the same,
 for instance $\tilde{Z}^u_L = \tilde{Z}^d_L$ and $\tilde{Z}^{e}_L = \tilde{Z}^{\nu}_L$.
This expressions allow us to calculate the weak mixing angles and a CP phase.

\subsection{Fermion sector of our model and results}
In the setup of our model, we introduce five bulk fermions
	$\Psi_{20},\Psi_{15},\Psi_{15'},\Psi_{6},\Psi_{6'}$
	and the corresponding mirror fermions shown in Table~\ref{Table1}.
Note that the number of the bulk fermions has been reduced from nine in \cite{MY2} to five, 
 which necessarily introduce generational mixings through the coupling of bulk fermions to boundary fermions.  
Lagrangian for the bulk and mirror fermions is given by
  \begin{eqnarray}
    \mathcal{L}_{\textmd{bulk+mirror}}
			= \sum_{a=20,\,15,\,15',\,6,\,6'}
				\left[\overline{\Psi}_a i\Gamma^MD_M\Psi_a
							+ \overline{\Tilde{\Psi}}_a i\Gamma^MD_M\Tilde{\Psi}_a
							+ \left(\frac{\lambda_a}{\pi R}\overline{\Psi}_a\Tilde{\Psi}_a
							+ \mbox{h.c.}\right)
				\right],
  \end{eqnarray}
	where the subscript $``a"$ denotes the $SU(6)$ representations of the bulk and mirror fermions.
The bulk masses between the bulk and the mirror fermions are normalized
	by $\pi R$ and expressed by the dimensionless parameter $\lambda_a$.
    \begin{table}[t]
        \centering
        \begin{tabular}{|c|c|} \hline
            bulk fermion $SU(6) \rightarrow SU(5)$ & mirror fermion \\ \hline
            $20^{(+,+)}=10\oplus10^*$ & $20^{(-,-)}$ \\ \hline
            $15^{(+,+)}=10\oplus5$ & $15^{(-,-)}$ \\ \hline
            $15'^{(+,-)}=10'\oplus5'$ & $15'^{(-,+)}$ \\ \hline
            $6^{(-,-)}=5\oplus1$ & $6^{(+,+)}$ \\ \hline
            $6'^{(+,+)}=5'\oplus1'$ & $6'^{(-,-)}$ \\ \hline
        \end{tabular}
        \vspace{-5pt}
        \caption{Representation of bulk fermions and the corresponding mirror fermions.
									$P_i$ are parity of bulk fermion for $\textbf{i}$ representation in $SU(6)$.
									$R$ in $R^{(+,+)}$ means an $SU(6)$ representation of the bulk fermion.
									$r_i$ in $r_1\oplus r_2$ are $SU(5)$ representations.}
        \label{Table1}
    \end{table}

The SM quarks and leptons for the first and the second generation
	are embedded into $SU(5)$ multiplets localized at $y=0$ boundary,
	which are two sets of decouplet, anti-quintet and singlet
	$\chi_{10},\, \chi_{5^*},\, \chi_1$.
On the other hand,
	those for the third generation are embedded into
	$SU(3)_C\times SU(2)_L\times U(1)_Y$ multiplets
	localized at $y=\pi R$ boundary.  
The reason why such a configuration of the SM fermions is adopted is
 to avoid massless SM quarks and leptons.
If three generations of the SM fermions are localized on $y=0$ boundary,
 we found that the rank of the mass matrices for the SM quarks and leptons at most two,
 which means that at least one massless quark or lepton is inevitable.
Therefore, Lagrangian for the SM fermions $\mathcal{L}_{\textmd{SM}}$ is expressed by
    \begin{eqnarray}
        \mathcal{L}_{\textmd{SM}}^{j=1,2} = \delta(y)[\Bar{\chi}^j_{10}i\Gamma^{\mu}D_{\mu}\chi^j_{10}
        +\Bar{\chi}^j_{5^*}i\Gamma^{\mu}D_{\mu}\chi^j_{5^*}+\Bar{\chi}^j_{1}i\Gamma^{\mu}D_{\mu}\chi^j_{1}], \nonumber \\
        \mathcal{L}_{\textmd{SM}}^{j=3} = \delta(y-\pi R)[\Bar{q}^3_Li\Gamma^{\mu}D_{\mu}q^3_L
        + \Bar{u}^3_Ri\Gamma^{\mu}D_{\mu}u^3_R + \Bar{d}^3_Ri\Gamma^{\mu}D_{\mu}d^3_R \nonumber \\
        +\Bar{l}^3_Li\Gamma^{\mu}D_{\mu}l^3_L +\Bar{e}^3_Ri\Gamma^{\mu}D_{\mu}e^3_R
        + \Bar{\nu}^3_Ri\Gamma^{\mu}D_{\mu}\nu^3_R].
    \end{eqnarray}
Here the superscript $``j"$ denotes the generation of the SM fermions, and $\Bar{\chi}_b\,(b=10, 5, 1)$ is Dirac conjugate of $\chi_b$.

    \begin{table}[t]
        \centering
        \begin{tabular}{|c|c|} \hline
					  bulk fermion $SU(5)\rightarrow SU(3)_C\times SU(2)_L\times U(1)_Y$ & SM fermion coupling to bulk \\ \hline
            $10 = Q_{20}(3,2)_{1/6}^{(+,+)}
									\oplus U_{20}^{\ast}(3^{\ast},1)_{-2/3}^{(+,-)}
									\oplus E^*_{20}(1,1)_{1}^{(+,-)}$
							& $q_L(3,2)_{1/6},~u_R^c(3^*,1)_{-2/3},~e_R^c(1,1)_{1}$\\ \hline
						$10^{\ast}
								= Q_{20}^{\ast}(3^{\ast},2)_{-1/6}^{(-,-)}
									\oplus U_{20}(3,1)_{2/3}^{(-,+)}
									\oplus E_{20}(1,1)_{-1}^{(-,+)}$
						  & $q^c_L(3^*,2)_{-1/6},~u_R(3,1)_{2/3},~e_R(1,1)_{-1}$\\ \hline
				\end{tabular}
			  \vspace{-10pt}
        \caption{$\textbf{20}$ bulk fermion and SM fermions.
									$r_{1,2}$ in $(r_1,~r_2)_a$ are $SU(3),~SU(2)$ representations in the SM, respectively.
									$a$ is $U(1)_Y$ charges.}
        \label{Table2}
\vspace{15pt}
        \centering
        \begin{tabular}{|c|c|} \hline
					  bulk fermion $SU(5)\rightarrow SU(3)_C\times SU(2)_L\times U(1)_Y$ & SM fermion coupling to bulk \\ \hline
            $10 = Q_{15}(3,2)_{1/6}^{(+,-)}
											\oplus U^*_{15}(3^*,1)_{-2/3}^{(+,+)}
											\oplus E^*_{15}(1,1)_{1}^{(+,+)}$
								& $q_L(3,2)_{1/6},~u_R^c(3^*,1)_{-2/3},~e_R^c(1,1)_{1}$\\ \hline
            $5 = D_{15}(3,1)_{-1/3}^{(-,+)}
										\oplus L^*_{15}(1,2)_{1/2}^{(-,-)}$
							& $d_R(3,1)_{-1/3},~l^c_L(1,2)_{1/2}$\\ \hline \hline
            bulk fermion $SU(5)\rightarrow SU(3)_C\times SU(2)_L\times U(1)_Y$ & SM fermion coupling to bulk \\ \hline
						$10' = Q_{15'}(3,2)_{1/6}^{(+,+)}
										\oplus U^*_{15'}(3^*,1)_{-2/3}^{(+,-)}
										\oplus E^*_{15'}(1,1)_{1}^{(+,-)}$
							& $q_L(3,2)_{1/6},~u_R^c(3^*,1)_{-2/3},~e_R^c(1,1)_{1}$\\ \hline
            $5' = D_{15'}(3,1)_{-1/3}^{(-,-)}
										\oplus L^*_{15'}(1,2)_{1/2}^{(-,+)}$
							& $d_R(3,1)_{-1/3},~l^c_L(1,2)_{1/2}$\\ \hline%
        \end{tabular}
        \vspace{-10pt}
        \caption{Upper (Lower) table shows $\textbf{15}$ ($\textbf{15}'$) bulk fermion and SM fermions.
									$r_{1,2}$ in $(r_1,~r_2)_a$ are $SU(3),~SU(2)$ representations in the SM, respectively.
									$a$ is $U(1)_Y$ charges.}
        \label{Table3}
\vspace{15pt}
        \centering
        \begin{tabular}{|c|c|} \hline
            bulk fermion $SU(5)\rightarrow SU(3)_C\times SU(2)_L\times U(1)_Y$ & SM fermion coupling to bulk \\ \hline
            $5 = D_{6}(3,1)_{-1/3}^{(-,+)}
									\oplus L^*_{6}(1,2)_{1/2}^{(-,-)}$
							& $d_R(3,1)_{-1/3},~l^c_L(1,2)_{1/2}$ \\ \hline
            $1 = N^*_{6}(1,1)_{0}^{(+,+)}$
							& $\nu^c_R(1,1)_{0}$\\ \hline \hline
            bulk fermion $SU(5)\rightarrow SU(3)_C\times SU(2)_L\times U(1)_Y$ & SM fermion coupling to bulk \\ \hline
            $5' = D_{6'}(3,1)_{-1/3}^{(+,-)}
									\oplus L^*_{6'}(1,2)_{1/2}^{(+,+)}$
							& $d_R(3,1)_{-1/3},~l^c_L(1,2)_{1/2}$ \\ \hline
            $1' = N^*_{6'}(1,1)_{0}^{(-,-)}$
							& $\nu^c_R(1,1)_{0}$\\ \hline
        \end{tabular}
        \vspace{-5pt}
        \caption{Upper (Lower) table shows $\textbf{6}$ ($\textbf{6}'$) bulk fermion and SM fermions.
									$r_{1,2}$ in $(r_1,~r_2)_a$ are $SU(3),~SU(2)$ representations in the SM, respectively.
									$a$ is $U(1)_Y$ charges.}
        \label{Table4}
    \end{table}

In order to realize the SM fermion masses,
	the boundary localized mass terms between the SM fermions localized at the boundaries and the bulk fermions
	are necessary.
To allow such localized mass terms,
	we have to choose appropriate $SU(6)$ representations for bulk fermions carefully.
Note that, for simplicity,
	the mirror fermions have no coupling to the SM fermions in this setup.
$\mathcal{L}_{\textmd{SM+bulk}}$ are boundary mass terms between the bulk fermions and the SM fermions,
	which are defined as follows.
  \begin{eqnarray}
        \mathcal{L}_{\textmd{SM+bulk}}^{j=1,2} &=& \delta(y)\sqrt{\frac{2}{\pi R}}[
        \epsilon^j_{20}(\Bar{\chi}^j_{10}\Psi_{10\subset20}+\Bar{\chi}^{j,c}_{10}\Psi_{10^*\subset20}) \nonumber \\
        && +\epsilon^j_{15}(\Bar{\chi}^j_{10}\Psi_{10\subset15}+\Bar{\chi}^{j,c}_{5^*}\Psi_{5\subset15})
        +\epsilon^j_{15'}(\Bar{\chi}^j_{10}\Psi_{10'\subset15'}+\Bar{\chi}^{j,c}_{5^*}\Psi_{5'\subset15'}) \nonumber \\
        && +\epsilon^j_{6}(\Bar{\chi}^{j,c}_{5^*}\Psi_{5\subset6}+\Bar{\chi}^j_{1}\Psi_{1\subset6})
        +\epsilon^j_{6'}(\Bar{\chi}^{j,c}_{5^*}\Psi_{5'\subset6'}+\Bar{\chi}^j_{1}\Psi_{1'\subset6'})+\mbox{h.c.}],
    \label{Lagrangian of SM + bulk for 1st and 2nd generation}
  \end{eqnarray}
  \begin{eqnarray}
        \mathcal{L}_{\textmd{SM+bulk}}^{j=3} &=& \delta(y-\pi R)\sqrt{\frac{2}{\pi R}}[
        \epsilon_{20e}(\Bar{e}^3_R E_{20}+\Bar{u}^3_R U_{20}) + \epsilon_{20q}\Bar{q}^3_L Q_{20} \nonumber \\
        && +\epsilon_{15u}\Bar{u}^{3,c}_R U^*_{15}+\epsilon_{15e}(\Bar{e}^{3,c}_R E^*_{15}+\Bar{l}^{3,c}_L L^*_{15})
        +\epsilon_{15'd} (\Bar{q}^3_L Q_{15'}+\Bar{d}^3_R D_{15'}) \nonumber \\
        && +\epsilon_{6\nu}(\Bar{l}^{3,c}_L L^*_6 + \Bar{\nu}^{3,c}_R N^*_{6})
        +\epsilon_{6'd} \Bar{d}^3_R D_{6'}+\mbox{h.c.}],
    \label{Lagrangian of SM + bulk for 3rd generation}
  \end{eqnarray}
where $\Psi_{M\subset N}$ is a bulk fermion for $M$ in $SU(5)$ representation and $N$ means $SU(6)$ representation.
$\epsilon$ are the strength of the mixing term between the bulk fermion and the SM fermion
	and should be complex numbers
	so that we can avoid a problem that the determinant of mass matrix in eq.~(\ref{eqmassM}) equals to zero. 
In other words, some SM fermions become massless state.
The decomposition of the introduced bulk fermions
 in the $\textbf{20}$, $\textbf{15}$ ($\textbf{15}'$), $\textbf{6}$ ($\textbf{6}'$) representations into the SM gauge group
 and the corresponding the SM fermions to be coupled on the boundary are summarized
 in Tables $\ref{Table2}$, $\ref{Table3}$, $\ref{Table4}$, respectively.

Based on the discussion above, our total Lagrangian for the fermions is given as follows:
    \begin{equation}
        \mathcal{L}_{\textmd{matter}} = \mathcal{L}_{\textmd{bulk+mirror}}
        + \mathcal{L}_{\textmd{SM}} + \mathcal{L}_{\textmd{SM+bulk}}.
    \end{equation}
%
%
%
%
%
Solving the exact KK spectrum of the bulk fermions from this Lagrangian is very hard task
 because of the complicated bulk and boundary system.
We assume in this paper that the physical mass induced for the boundary fields
	is much smaller than the masses of the bulk fields \cite{SSS}.
This is reasonable since the compactification scale and the bulk mass mainly determining the KK mass spectrum of the bulk fields
 is larger than the mass for the boundary fields whose typical scale is given by the Higgs VEV.
In this case, the effects of the mixing on the spectrum for the bulk fields can be negligible and the spectrum
	$m_n^2 = (\frac{\lambda}{\pi R})^2 + m_n{(q\alpha)^2}$ is a good approximation \cite{SSS}.

Thanks to the reduction of bulk fermions, we can reproduce the quark and lepton mixing angles in addition to the SM fermion masses. For example, the discussion in section 3.1 allows us to obtain up-type quark mass $M^u$ can be written in the following form
\begin{equation}
\begin{aligned}
    M^u = m_W\sqrt{1+c} \left(
    \begin{array}{ccc}
        2\epsilon^1_{20} \epsilon^{1*}_{20} \Lambda_{11} & (\epsilon^1_{20} \epsilon^{2*}_{20} + \epsilon^2_{20} \epsilon^{1*}_{20}) \Lambda_{12} & (\epsilon^1_{20} \epsilon^*_{20u} + \epsilon_{20u} \epsilon^{1*}_{20}) \Lambda_{13} \\
        (\epsilon^2_{20} \epsilon^{1*}_{20} + \epsilon^1_{20} \epsilon^{2*}_{20}) \Lambda_{21} & 2\epsilon^2_{20} \epsilon^{2*}_{20} \Lambda_{22} & (\epsilon^2_{20} \epsilon^*_{20u} + \epsilon_{20u}\epsilon^{2*}_{20}) \Lambda_{23} \\
        (\epsilon_{20q} \epsilon^{1*}_{20} + \epsilon^{1}_{20} \epsilon^*_{20q}) \Lambda_{31} & (\epsilon_{20q} \epsilon^{2*}_20 + \epsilon^{2}_20 \epsilon^*_{20q}) \Lambda_{32} & (\epsilon_{20q} \epsilon_{20u}^* + \epsilon_{20u} \epsilon^*_{20q}) \Lambda_{33}
    \end{array}\right)
\end{aligned}
\end{equation}
where $c=c_1 + c_2$, and $\epsilon^i_{20}\,(i=1,2), \epsilon_{20q}, \epsilon_{20u}$ are the input parameters that appeared in eq. (\ref{Lagrangian of SM + bulk for 1st and 2nd generation}) and eq. (\ref{Lagrangian of SM + bulk for 3rd generation}). $\Lambda_{ij}\,(i,j=1,2,3)$ is defined as $\Lambda_{ij} = \coth^2|\lambda_{20}|+\tanh^2|\lambda_{20}|-2\,(i,j=1,2)$, $\Lambda_{i3}=\lambda_{3i}=\coth{|\lambda_{20}|} \textmd{cosec}|\lambda_{20}|\,(i=1,2)$, and $\Lambda_{33}=\coth^2|\lambda_{20}|-1$. Mass matrices of other type quark and lepton are given in a similar way. After following the way discussed in section 3.2, we can also calculate the flavor mixing angles and a CP phase and fit them to their experimental values. This is because we have now changed the way the bulk fermions couple to the SM fermions in this paper while each type of bulk fermions was introduced per generation so that the mass matrices were essentially diagonal in the previous model.

Finally, we have found allowed parameter sets to reproduce the SM fermion masses and mixing
 and some sample data sets depending on the parameter of the localized gauge kinetic terms $c$
 are shown in Tables \ref{Table5}, \ref{Table6}, \ref{Table7}. Note that we use in our analysis the experimental data
 and one of standard conventions for CKM and PMNS matrix shown in Particle Data Group \cite{PDG}.
In the analysis for neutrino sector, the normal hierarchy is assumed although it is not essential.
As can be seen from the Tables \ref{Table5}, \ref{Table6},
 our results are in almost good agreement with the experimental data.
This is very remarkable result since the generation mixings in the bulk are newly introduced
 resulting from the reduction of the number of bulk fermions,
 which makes, in particular, reproducing the quark and lepton mixing angles highly nontrivial.
Our model discussed in this paper turned out to be a good starting point for constructing a realistic model of GGHU.

    \begin{table}[b]
        \centering
        \begin{tabular}{|c|c||c|c|c|c|} \hline
            $1/R$ & $c$ & $m_u$ & $m_c$ & $m_t$ & \\ \hline
            $10\mbox{TeV}$ & 80 & 2.163 MeV & 1.217 GeV & 166.294 GeV & \\ \hline
            $10\mbox{TeV}$ & 90 & 2.320 MeV & 1.229 GeV & 167.931 GeV & \\ \hline
            $15\mbox{TeV}$ & 80 & 2.316 MeV & 1.214 GeV & 165.300 GeV & \\ \hline
						$15\mbox{TeV}$ & 90 & 2.156 MeV & 1.225 GeV & 166.89 GeV & \\ \hline
            Data & & $2.16^{+0.49}_{-0.26}$ MeV & $1.27\pm0.02$ GeV & $172\pm0.30$ GeV & \\ \hline \hline
            $1/R$ & $c$ & $m_d$ & $m_s$ & $m_b$ & \\ \hline
            $10\mbox{TeV}$ & 80 & 5.583 MeV & 75.7 MeV & 4.155 GeV & \\ \hline
            $10\mbox{TeV}$ & 90 & 5.505 MeV & 75.8 MeV & 4.321 GeV & \\ \hline
            $15\mbox{TeV}$ & 80 & 5.522 MeV & 75.5 MeV & 4.201 GeV & \\ \hline
						$15\mbox{TeV}$ & 90 & 5.545 MeV & 75.1 MeV & 4.183 GeV & \\ \hline
            Data & & $4.67^{+0.48}_{-0.17}$ MeV & $93^{+11}_{-5}$ MeV & $4.18^{+0.13}_{-0.02}$ GeV & \\ \hline \hline
            $1/R$ & $c$ & $\sin\theta_{12}$ & $\sin\theta_{13}$ & $\sin\theta_{23}$ & $\delta$ \\ \hline
            $10\mbox{TeV}$ & 80 & 0.191797 & 0.003537 & 0.041430 & 1.1560 \\ \hline
            $10\mbox{TeV}$ & 90 & 0.195857 & 0.003510 & 0.039893 & 1.2424 \\ \hline
            $15\mbox{TeV}$ & 80 & 0.190839 & 0.003556 & 0.041459 & 1.1831 \\ \hline
						$15\mbox{TeV}$ & 90 & 0.192085 & 0.003518 & 0.040088 & 1.1750 \\ \hline
            Data & & $0.22650\pm0.00048$ & $0.00361^{+0.00011}_{-0.00009}$ & $0.04053^{+0.00083}_{-0.00061}$ & $1.196^{+0.045}_{-0.043}$  \\ \hline
        \end{tabular}
        \caption{Our results of parameter fitting in quark sector for some parameters $1/R$ and $c=c_1+c_2$. The up, down, and strange quark masses are the $\overline{\textmd{MS}}$ masses at the scale $\mu$ = 2 GeV. The charm and bottom quark masses are the $\overline{\textmd{MS}}$ masses renormalized at the $\overline{\textmd{MS}}$ mass, i.e. $\Bar{m} = \Bar{m}(\mu = \Bar{m})$. The top quark mass is extracted from direct measurements.}
        \label{Table5}
    \end{table}

    \begin{table}[b]
        \centering
        \begin{tabular}{|c|c||c|c|c|c|} \hline
            $1/R$ & $c$ & $m_e$ & $m_{\mu}$ & $m_{\tau}$ \\ \hline
            $10\mbox{TeV}$ & 80 & 0.5136 MeV & 98.750 MeV & 1687.12 MeV \\ \hline
            $10\mbox{TeV}$ & 90 & 0.5140 MeV & 98.188 MeV & 1689.56 MeV \\ \hline
            $15\mbox{TeV}$ & 80 & 0.5135 MeV & 98.776 MeV & 1695.46 MeV \\ \hline
						$15\mbox{TeV}$ & 90 & 0.5139 MeV & 98.610 MeV & 1687.59 MeV \\ \hline
            Data & & 0.5109989461(31) MeV & 105.6583745(24) MeV & 1776.86(12) MeV\\ \hline \hline
            $1/R$ & $c$ & $\Delta m_{21}^2$ & $\Delta m_{32}^2$ (Normal) & $\delta$ \\ \hline
            $10\mbox{TeV}$ & 80 & $7.7306\times10^{-5}\,\textmd{eV}^2$ & $2.4524\times10^{-3}\,\textmd{eV}^2$ & 1.539$\pi$ rad \\ \hline
            $10\mbox{TeV}$ & 90 & $7.7087\times10^{-5}\,\textmd{eV}^2$ & $2.4367\times10^{-3}\,\textmd{eV}^2$ & 1.536$\pi$ rad \\ \hline
            $15\mbox{TeV}$ & 80 & $7.8054\times10^{-5}\,\textmd{eV}^2$ & $2.3895\times10^{-3}\,\textmd{eV}^2$ & 1.531$\pi$ rad \\ \hline
						$15\mbox{TeV}$ & 90 & $7.6544\times10^{-5}\,\textmd{eV}^2$ & $2.4577\times10^{-3}\,\textmd{eV}^2$ & 1.536$\pi$ rad \\ \hline
            Data & & $(7.53\pm0.18)\times10^{-5}\,\textmd{eV}^2$ & $(2.453\pm0.033)\times10^{-3}\,\textmd{eV}^2$ & $1.36^{+0.20}_{-0.16}\pi$ rad
            \\ \hline \hline
            $1/R$ & $c$ & $\sin^2\theta_{12}$ & $\sin^2\theta_{13}$ & $\sin^2\theta_{23}$ (Normal) \\ \hline
            $10\mbox{TeV}$ & 80 & 0.3313 & $2.240\times10^{-2}$ & 0.5161 \\ \hline
            $10\mbox{TeV}$ & 90 & 0.3294 & $2.155\times10^{-2}$ & 0.5187 \\ \hline
            $15\mbox{TeV}$ & 80 & 0.3505 & $2.094\times10^{-2}$ & 0.5069 \\ \hline
						$15\mbox{TeV}$ & 90 & 0.3308 & $2.123\times10^{-2}$ & 0.5161 \\ \hline
            Data & & $0.307\pm0.013$ & $(2.20\pm0.07)\times10^{-2}$ & $0.546\pm0.021$  \\ \hline
        \end{tabular}
        \caption{Our results of parameter fitting in lepton sector for some parameters $1/R$ and $c=c_1+c_2$.
        In neutrino sector, normal hierarchy is assumed.}
        \label{Table6}
    \end{table}

    \begin{table}[b]
        \centering
        \begin{tabular}{|c|c|c|c|c|c|c|} \hline
            $1/R$ & $c$ & $\lambda_{20}$ & $\lambda_{15}$ & $\lambda_{15'}$ & $\lambda_{6}$ & $\lambda_{6'}$ \\ \hline
            $10\mbox{TeV}$ & 80 & 0.697103 & 0.379299 & 1.86668 & 13.404 & 12.0452 \\ \hline
            $10\mbox{TeV}$ & 90 & 0.707565 & 0.382020 & 1.87516 & 13.383 & 12.0376 \\ \hline
            $15\mbox{TeV}$ & 80 & 0.727571 & 0.421544 & 1.88562 & 13.390 & 11.0374 \\ \hline
			$15\mbox{TeV}$ & 90 & 0.737359 & 0.428899 & 1.89630 & 13.355 & 11.0374 \\ \hline \hline
            $1/R$ & $c$ & $|\epsilon^1_{20}|$ & $|\epsilon^1_{15}|$ & $|\epsilon^1_{15'}|$ & $|\epsilon^1_{6}|$ & $|\epsilon^1_{6'}|$ \\ \hline
            $10\mbox{TeV}$ & 80 & 1 & 0.0618938 & 0.398107 & 0.000724436 & 0.0204174 \\ \hline
            $10\mbox{TeV}$ & 90 & 1.00005 & 0.0624081 & 0.398146 & 0.000668344 & 0.0197242  \\ \hline
            $15\mbox{TeV}$ & 80 & 0.99960 & 0.0620899 & 0.398107 & 0.0000316228 & 0.00668344  \\ \hline
			$15\mbox{TeV}$ & 90 & 0.999994 & 0.062593 & 0.398095 & 0.0000295121 & 0.00653131 \\ \hline \hline
            $1/R$ & $c$ & $|\epsilon^2_{20}|$ & $|\epsilon^2_{15}|$ & $|\epsilon^2_{15'}|$ & $|\epsilon^2_{6}|$ & $|\epsilon^2_{6'}|$ \\ \hline
            $10\mbox{TeV}$ & 80 & 0.190546 & 0.0141508 & 0.0794328 & 0.00042658 & 0.0101158 \\ \hline
            $10\mbox{TeV}$ & 90 & 0.190545 & 0.0141773 & 0.0794333 & 0.000380189 & 0.00988553 \\ \hline
            $15\mbox{TeV}$ & 80 & 0.190546 & 0.0141593 & 0.0794328 & 0.0000285102 & 0.0049545 \\ \hline
			$15\mbox{TeV}$ & 90 & 0.190546 & 0.0141970 & 0.0794331 & 0.0000266073 & 0.00484172 \\ \hline \hline
            $1/R$ & $c$ & $\theta^2_{20}$ & $\theta^2_{15}$ & $\theta^2_{15'}$ & $\theta^2_{6}$ & $\theta^2_{6'}$ \\ \hline
            $10\mbox{TeV}$ & 80 & 0.00736813 & 0.258731 & 0.0703712 & 5.956450 & 3.78311 \\ \hline
            $10\mbox{TeV}$ & 90 & 0.00744981 & 0.253490 & 0.0710904 & 5.956450 & 3.78311 \\ \hline
            $15\mbox{TeV}$ & 80 & 0.00775807 & 0.268290 & 0.0731497 & 0.712979 & 4.94097 \\ \hline
			$15\mbox{TeV}$ & 90 & 0.00774413 & 0.263562 & 0.0743995 & 0.752089 & 4.94097 \\ \hline \hline
            $1/R$ & $c$ & $|\epsilon_{20e}|$ & $|\epsilon_{15u}|$ & $|\epsilon_{15'd}|$ & $|\epsilon_{6\nu}|$ & $|\epsilon_{6'd}|$ \\ \hline
            $10\mbox{TeV}$ & 80 & 0.0542386 & 0.000873495 & 0.00877701 & 0.00001 & 0.00001 \\ \hline
            $10\mbox{TeV}$ & 90 & 0.0522386 & 0.000847602 & 0.00849188 & 0.0000101158 & 0.00001 \\ \hline
            $15\mbox{TeV}$ & 80 & 0.0542109 & 0.000878758 & 0.00888508 & 0.000213796 & 0.00001 \\ \hline
			$15\mbox{TeV}$ & 90 & 0.0519660 & 0.000853353 & 0.00858460 & 0.000211349 & 0.00001 \\ \hline \hline
            $1/R$ & $c$ & $\theta_{20e}$ & $\theta_{15u}$ & $\theta_{15'd}$ & $\theta_{6\nu}$ & $\theta_{6'd}$ \\ \hline
            $10\mbox{TeV}$ & 80 & 4.25155 & 4.14707 & 4.21375 & 3.18033 & 0.00001 \\ \hline
            $10\mbox{TeV}$ & 90 & 4.22678 & 4.16561 & 4.21489 & 3.18033 & 0.00001  \\ \hline
            $15\mbox{TeV}$ & 80 & 4.28915 & 4.11134 & 4.22385 & 4.56062 & 0.00001 \\ \hline
			$15\mbox{TeV}$ & 90 & 4.28632 & 4.12957 & 4.2136 & 4.56062 & 0.00001 \\ \hline \hline
            $1/R$ & $c$ & $|\epsilon_{20q}|$ & $|\epsilon_{15e}|$ & $\theta_{20q}$ & $\theta_{15e}$ & \\ \hline
            $10\mbox{TeV}$ & 80 & 0.0113599 & 0.00122572 & 2.15576 & 2.16323 & \\ \hline
            $10\mbox{TeV}$ & 90 & 0.0109116 & 0.00196532 & 2.13719 & 2.59540 & \\ \hline
            $15\mbox{TeV}$ & 80 & 0.0115976 & 0.00123553 & 2.19391 & 2.57235 & \\ \hline
			$15\mbox{TeV}$ & 90 & 0.0114167 & 0.00140787 & 2.17591 & 2.65110 & \\ \hline
        \end{tabular}
        \caption{Input parameters of the parameter fitting where $c=c_1+c_2$. 
        $\lambda_a\,(a=20,15,15',6,6')$ are the bulk masses 
        between the bulk and the mirror fermions which are normalized by $\pi R$. 
        $\epsilon$ is the strength of the mixing term between the bulk fermion and the SM fermion, 
        $|\epsilon|$ is its absolute value, and $\theta$ is a phase of the corresponding $\epsilon$. 
        Only $\epsilon^1_a\,(a=20,15,15',6,6')$ can be taken as a real number without loss of generality. 
        This is because we have not shown $\theta^1_a\,(a=20,15,15',6,6')$ in this table.}
        \label{Table7}
    \end{table}

\clearpage
\section{Conclusions}
In this paper, we have discussed the fermion mass hierarchy and mixing in $SU(6)$ GGHU with localized gauge kinetic terms.
The SM fermions are introduced on the boundaries.
We also introduced massive bulk fermions in three types of $SU(6)$ representations coupling to the SM fermions on the boundaries.
The number of them has been reduced in order to achieve generation mixings of quarks and leptons,
 which greatly changed the coupling of the bulk fermions to the SM fermions on the boundaries
 and additional generation mixings in the bulk sector appeared.
This feature makes our analysis on the SM fermion masses and mixing angles highly complicated and nontrivial.
We have shown that the SM fermion masses and mixing can be almostly reproduced by mild tuning of bulk masses
 and the parameters of the localized gauge kinetic terms.
Some parameter sets of our results are listed.
Our model discussed in this paper turned out to be a good starting point for constructing a realistic model of GGHU.

As remained for our future work,
 it is important to calculate the effective potential for the Higgs field and study whether the electroweak symmetry breaking correctly occurs.
Since the Higgs field is originally a gauge field, the potential is generated at one-loop by Coleman-Weinberg mechanism.
It is not easy to obtain the observed Higgs mass 125 GeV because the effects of localized gauge kinetic terms
 enhance the compactification scale and also Higgs boson mass.
In viewpoint of the gauge coupling unification, we have little room for introducing extra bulk fields to adjust Higgs mass,
 which also makes the analysis of the electroweak symmetry breaking difficult.
It is possible to introduce Majorana neutrino on the boundary,
 which relaxes the constraint for the bulk mass parameters, in order to obtain the Higgs boson mass.
We would investigate the electroweak symmetry breaking and Higgs boson mass along this line.

\subsection*{Acknowledgments}
This work was supported by JST SPRING, Grant Number JPMJSP2139 (Y.Y.).

\vspace*{5mm}

\appendix

\section{The propagator of the bulk and mirror fermion}
The bulk and mirror fermions $\Psi(x,y), \tilde{\Psi}(x,y)$ live in the five dimension and
	their KK decomposition is given by
	\begin{equation}
  	\begin{aligned}
    	\Psi(x,y) &= \sum_n f_n(y) \chi_n(x), \\
    	\tilde{\Psi}(x,y) &= \sum_n f_n(y) \tilde{\chi}_n(x),
  	\end{aligned}
	\end{equation}
where $\chi_n(x), \tilde{\chi}_n(x)$ are 4D fields and $f_n(y)$ is a mode function.
The quadratic terms for $\chi_n(x), \tilde{\chi}_n(x)$ can be expressed in momentum space
 by using the above KK decomposition as
	\begin{equation}
  	(\chi_n^{\dag}, \tilde{\chi}_n^{\dag})
  	\left(
  	\begin{array}{cc}
    	\Slash{p} -m_n & M \\
    	M & \Slash{p} +m_n \\
  	\end{array}
  	\right)
  	\left(
  	\begin{aligned}
    	\chi_n\\
    	\tilde{\chi}_n\\
  	\end{aligned}
  	\right),
	\end{equation}
which leads to the propagator in Minkowski spacetime
	\begin{equation}
  	\Delta_{\chi}
  	=
  	\frac{i}{p^2-m_n^2-M^2}
  	\left(
  	\begin{array}{cc}
    	\Slash{p} +m_n & - M \\
    	-M & \Slash{p} - m_n \\
  	\end{array}
  	\right),
	\end{equation}
	or in Euclidean spacetime
	\begin{equation}
  	\Delta_{\chi}^E =i \Delta_{\chi}
  	=
  	\frac{1}{p^2_E +m_n^2 +M^2}
  	\left(
  	\begin{array}{cc}
    	i\Slash{p}_E -m_n &  M \\
     	M & i\Slash{p} + m_n \\
  	\end{array}
  	\right).
	\end{equation}

\section{Summation of the KK mode}
In order to evaluate Eq.(\ref{eq1}), we need to calculate the summation of KK mode contributions.
The summation of KK mode contributions can be rewritten as,
\begin{equation}
  \begin{aligned}
    f_{\delta}^{(T)}(x ,q\alpha)
    &= \sum_n \frac{(-1)^{\delta n}}{x +i\pi (n+\nu_T+\alpha)}\\
    &=\left\{
    \begin{array}{ll}
      \coth(x +i \pi \alpha) & :\delta=0, T=+1, (\nu_T=0)\\
      \sinh(x +i \pi \alpha)^{-1} & :\delta=1, T=+1, (\nu_T=0)\\
      \tanh(x +i \pi \alpha) & :\delta=0, T=-1, (\nu_T=1/2)\\
      \cosh(x +i \pi \alpha)^{-1} & :\delta=-1, T=-1, (\nu_T=1/2)\\
    \end{array}\right. ,
  \end{aligned}
\end{equation}
where 
$\delta=0(1)$ for the case where the left-handed and right-handed boundary fermions are localized on the same (opposite) boundary.
$T$ is the periodcity for bulk and mirror fermion.
The factor $n+ \nu_T +q\alpha$ in the denominator comes from the KK mass spectrum
\begin{equation}
  m_n = \frac{n+ \nu_T +q\alpha}{R}
\end{equation}
where $\alpha$ is a dimensionless parameter corresponding to VEV of Higgs field.

Furthermore, it is useful to express the real and imaginary part of these function.
\begin{equation}
  \begin{aligned}
    \mbox{Re}f_{\delta}^{(T)}(x ,q\alpha)
     &=\frac{1}{2}\sum_n
        \left(\frac{(-1)^{\delta n}}{x +i\pi (n+\nu_T+\alpha)}
              +\frac{(-1)^{\delta n}}{x -i\pi (n+\nu_T+\alpha)}\right)\\
    &=\sum_n \frac{x(-1)^{ \delta n}}{x^2 +\pi^2 (n+\nu_T+\alpha)^2},\\
  \end{aligned}
\end{equation}

\begin{equation}
  \begin{aligned}
    \mbox{Im}f_{\delta}^{(T)}(x ,q\alpha)
     &=\frac{1}{2i}\sum_n
        \left(\frac{(-1)^{\delta n}}{x +i\pi (n+\nu_T+\alpha)}
              -\frac{(-1)^{\delta n}}{x -i\pi (n+\nu_T+\alpha)}\right)\\
    &=\sum_n \frac{-  (-1)^{ \delta n} \pi (n+\nu_T+\alpha)}{x^2 +\pi^2 (n+\nu_T+\alpha)^2},\\
  \end{aligned}
\end{equation}
therefore we obtain
\begin{equation}
	\begin{aligned}
		\sum_n \frac{(-1)^{ \delta n}}{x^2 + (\pi R m_n)^2}
	 &= \frac{1}{x} \mbox{Re}f_{\delta}^{(T)}(x ,q\alpha), \\
	  \sum_n \frac{(-1)^{ \delta n} m_n}{x^2 +(\pi R m_n)^2}
 	 &=  -\frac{1}{\pi R}\mbox{Im}f_{\delta}^{(T)}(x ,q\alpha). \\
	\end{aligned}
\end{equation}



\vspace*{10mm}

\begin{thebibliography}{99}

\bibitem{GHU}
N.S.~Manton, Nucl. Phys. B \textbf{158}, 141 (1979);
D.B.~Fairlie, Phys. Lett. B \textbf{82}, 97 (1979), J. Phys. G \textbf{5}, L55 (1979);
Y. Hosotani, Phys. Lett. B \textbf{126}, 309 (1983), Phys. Lett. B \textbf{129}, 193 (1983), Annals Phys. \textbf{190}, 233 (1989).

\bibitem{1loop}
H. Hatanaka, T. Inami and C.S. Lim, Mod. Phys. Lett. A \textbf{13}, 2601 (1998);
I. Antoniadis, K. Benakli and M. Quiros, New J. Phys. \textbf{3}, 20 (2001);
G. von Gersdorff, N. Irges and M. Quiros, Nucl. Phys. B \textbf{635}, 127 (2002);
R. Contino, Y. Nomura and A. Pomarol, Nucl. Phys. B \textbf{671}, 148 (2003);
C.S. Lim, N. Maru and K. Hasegawa, J. Phys. Soc. Jap. \textbf{77}, 074101 (2008).

\bibitem{2loop}
N. Maru and T. Yamashita, Nucl. Phys. B \textbf{754}, 127 (2006);
Y. Hosotani, N. Maru, K. Takenaga and T. Yamashita, Prog. Thoer. Phys. \textbf{118}, 1053 (2007);











\bibitem{LM}
C.S. Lim and N. Maru, Phys. Lett. B \textbf{653}, 320 (2007).

\bibitem{others}
G. Burdman and Y. Nomura, Nucl. Phys. B \textbf{656}, 3 (2003);
N. Haba, Y. Hosotani, Y. Kawamura and T. Yamashita, Phys. Rev. D \textbf{70}, 015010 (2004);
K. Kojima, K. Takenaga and T. Yamashita, Phys. Rev. D \textbf{84}, 051701 (2011); Phys. Rev. D \textbf{95}, no. 1, 015021 (2017);
JHEP \textbf{1706}, 018 (2017);
Y. Hosotani and N. Yamatsu, PTEP \textbf{2015}, 111B01 (2015);
A. Furui, Y. Hosotani and N. Yamatsu, PTEP \textbf{2016}, no. 9, 093B01 (2016);
Y. Hosotani and N. Yamatsu, PTEP \textbf{2017}, no. 9, 091B01 (2017); PTEP \textbf{2018}, no. 2, 023B05 (2018);
S. Funatsu, H. Hatanaka, Y. Hosotani, Y. Orikasa and N. Yamatsu, Phys. Rev. D \textbf{99}, no. 9, 095010 (2019), Phys. Rev. D \textbf{102}, no.1, 015005 (2020); A.~Angelescu, A.~Bally, S.~Blasi and F.~Goertz, Phys. Rev. D \textbf{105}, no.3, 035026 (2022); A.~Angelescu, A.~Bally, F.~Goertz and S.~Weber, arXiv:2208.13782 [hep-ph].

\bibitem{SSS}
C.A. Scrucca, M. Serone and L. Silvestrini, Nucl. Phys. B \textbf{669}, 128 (2003).

\bibitem{CGM}
C. Csaki, C. Grojean and H. Murayama, Phys. Rev. D \textbf{67}, 085012 (2003).


\bibitem{MY1}
N. Maru and Y. Yatagai, PTEP \textbf{2019}, no. 8, 083B03 (2019); arXiv:1903.08359 [hep-ph].

\bibitem{CTW}
M. Carena, T.M.P. Tait and C.E.M. Wagner, Acta Phys. Polon. B \textbf{33}, 2355 (2002).

\bibitem{MY2}
N. Maru and Y. Yatagai, Eur. Phys. J. C, \textbf{80}(10):933 (2020); arXiv:1911.03465 [hep-ph].



\bibitem{Kawamura}
Y. Kawamura, Prog. Theor. Phys. \textbf{105}, 999 (2001).


\bibitem{PDG}
P.A. Zyla \textit{et al}. (Particle Data Group), Prog. Theor. Exp. Phys. \textbf{2020}, 083C01 (2020).


\end{thebibliography}
\end{document}